\documentclass[notitlepage,superscriptaddress]{revtex4-1}

\usepackage{amsfonts,amssymb,amsmath}
\usepackage[]{graphics,graphicx,epsfig}
\usepackage{amsthm}

\def\identity{\leavevmode\hbox{\small1\kern-3.8pt\normalsize1}}

\renewcommand{\epsilon}{\varepsilon}

\usepackage{color}

\begin{document}

\title{In-vivo biomagnetic characterisation of the American cockroach}

\author{Ling-Jun Kong}
\affiliation{School of Physical and Mathematical Sciences, Nanyang Technological University, 637371 Singapore}
\affiliation{MOE Key Laboratory of Weak Light Nonlinear Photonics and School of Physics, Nankai University, Tianjin 300071, China}

\author{Herbert Crepaz}
\affiliation{School of Physical and Mathematical Sciences, Nanyang Technological University, 637371 Singapore}
\affiliation{Centre for Quantum Technologies, National University of Singapore, 117543 Singapore}

\author{Agnieszka G\'orecka}
\affiliation{School of Physical and Mathematical Sciences, Nanyang Technological University, 637371 Singapore}
\affiliation{School of Physics and Astronomy, Monash University, Victoria 3800, Australia}

\author{Aleksandra Urbanek}
\affiliation{Department of Invertebrate Zoology and Parasitology, University of Gda\'nsk, 80-308 Poland}

\author{Rainer Dumke}
\affiliation{School of Physical and Mathematical Sciences, Nanyang Technological University, 637371 Singapore}
\affiliation{Centre for Quantum Technologies, National University of Singapore, 117543 Singapore}

\author{Tomasz Paterek}
\affiliation{School of Physical and Mathematical Sciences, Nanyang Technological University, 637371 Singapore}
\affiliation{Centre for Quantum Technologies, National University of Singapore, 117543 Singapore}

\begin{abstract}
We present a quantitative method, utilising a highly sensitive quantum sensor, that extends applicability of magnetorelaxometry to biological samples at physiological temperature.
The observed magnetic fields allow for non-invasive determination of physical properties of magnetic materials and their surrounding environment inside the specimen.
The method is applied to American cockroaches and reveals magnetic deposits with strikingly different behaviour in alive and dead insects.
We discuss consequences of this finding to cockroach magneto-reception. To our knowledge, this work represents the first characterisation of the magnetisation dynamics in live insects and helps to connect results from behavioural experiments on insects in magnetic fields with characterisation of magnetic materials in their corpses.
\end{abstract}

\maketitle

\section{Introduction}

Many species are capable of perceiving the world through senses inaccessible to humans.
Polarisation vision of marine species~\cite{Cronin.2003b} or magnetic field detection by migratory birds~\cite{Wiltschko.2005b} being two well-known examples.
Magneto-reception is in fact common to a wide range of organisms, ranging from bacteria to higher vertebrates, and has evolved to a fine-tuned sensory system that maybe even takes advantage of quantum coherence~\cite{Lambert.2012}. Insights into magneto-reception mechanisms and biomagnetism, magnetic fields that originate in biological systems, not only allow us to understand better different ways of visualising the world but may also find applications in improved man-made sensors inspired by their biological counterparts.

Several behavioural experiments have demonstrated that cockroaches and other insects are capable of magneto-reception, see e.g.~\cite{Vacha.2006b,Vacha.2008b,Vacha.2009b,vacha.pnas,vacha2010.quadrupodal,Camlitepe.2005,Wajnberg.review,BEE_ERRORS,BEE_DEMAG,BEE_TRAIN,BEE_TRAIN2,BEE_WEAK,BEE_WEAK2,BEE_PATTERN}.
Most of the behavioural experiments on cockroaches show their ability to perceive changes of an Earth-strength magnetic field that is rotated with a period of $10$ minutes.
A different set of experiments found and characterised magnetic particles in insect corpses, see e.g.~\cite{Gould.1978,Kuterbach.1986,AcostaAvalos.1999,Alves.2004,wajnberg.biometals,Maher.1998b,Pan.2016,MONARCH}.
Naturally one asks if these magnetic particles contribute to magneto-reception.
One route to answer this question is to develop techniques for tracking dynamics and quantifying the extent of magnetic materials inside a living insect.

\begin{figure}[!b]
\centering
\includegraphics{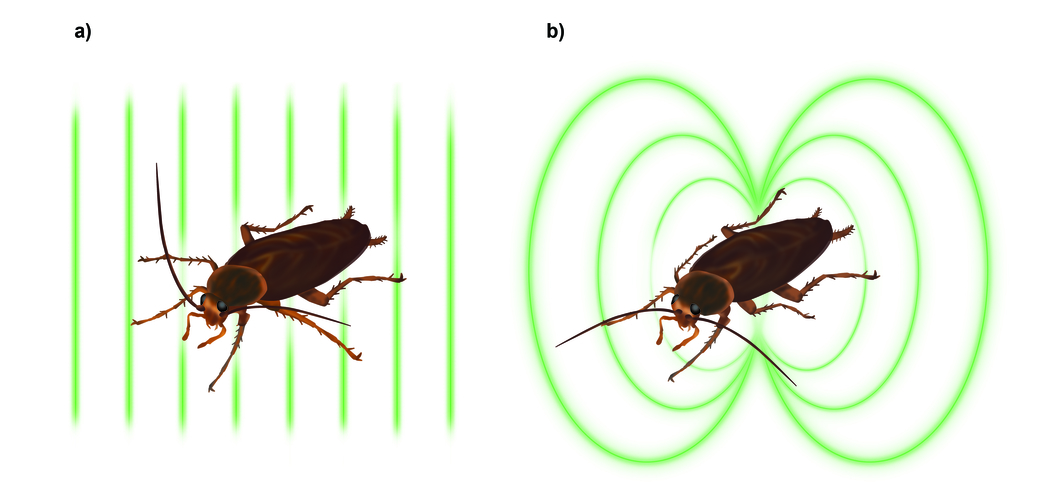} 
\caption{ {\small Sketch of the experiment.
(a) American cockroaches were placed in a strong magnetic field aligned perpendicular to the thorax as illustrated by the green lines.
Using an atomic magnetometer we monitored the dynamics of the magnetic field generated by the magnetised insects.
(b) The magnetic field is very close to the field of magnetic dipole normal to the thorax.}
Published with permission from T. Yeo.}
\label{fig_roach}
\end{figure}

Here we demonstrate a non-invasive method for magnetic field measurements taking advantage of the high precision of atomic magnetometer~\cite{Budker.}.
It is a form of magnetorelaxometry (MRX)~\cite{magnetorelaxometry} where the magnetic materials to be characterised could be present inside a living organism.
The method is applied to study magnetic fields generated by magnetised American cockroaches (\emph{Periplaneta americana}).
Our study reveals presence of magnetic materials in their bodies exhibiting distinct dynamics in alive and dead cockroaches.
After magnetisation of alive insects, we observe exponential magnetic field decay to a remnant value, with a decay time of $50 \pm 28$ minutes.
In contradistinction, an average demagnetisation of dead cockroaches displays a much longer decay time of $47.5 \pm 28.9$ hours.

This clear difference in magnetic field decay is explained by Brownian rotations of magnetic materials in different viscosity environments.
Fits of this model to the measured data reveal magnetic particles embedded in an environment which experiences a two orders of magnitude viscosity increment between alive and dead cockroaches.
Under the assumption that there were no external magnetic particles on measured insects we find sub-micron sized magnetic deposits and a glassy environment of high viscosity.
Such deposits would require several hours to align within an Earth-strength magnetic field.
Thus, they cannot be responsible for the observed cockroach magneto-reception in rotating magnetic fields with a period of $10$ minutes~\cite{Vacha.2006b,Vacha.2008b,Vacha.2009b,vacha.pnas}.
This provides further support to additional working mechanism behind magneto-reception, e.g. the radical-pair mechanism~\cite{Vacha.2009b,vacha.pnas}.

\section{Results}

\begin{figure}[!b]
\includegraphics[scale=0.23]{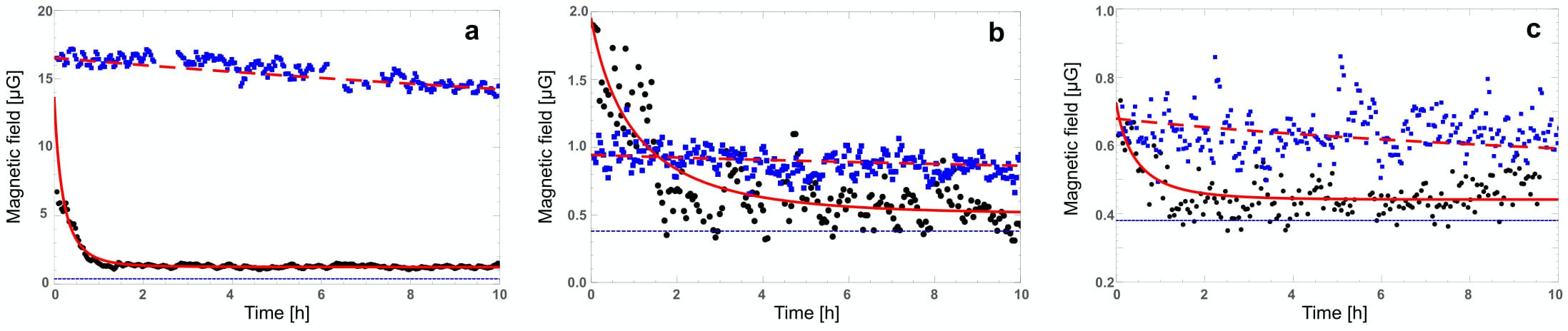}
\caption{{\small Magnetic field decay from magnetised American cockroaches.
Black dots show the measured time dependence of the magnetic field for alive cockroaches and blue squares show this dependence for the dead ones.
Different panels present exemplary data for different insects.
They were chosen to show the typical sets where the initial magnetisation of alive cockroach can be higher, similar or lower than the initial magnetisation of the dead one. 
This is captured in our model as well.
Altogether we conducted $15$ measurements lasting longer than $10$ hours each and additionally more than $10$ shorter measurements.
The thick red lines represent simulation of our model, fitting the data: solid for alive cockroaches and dashed for dead ones.
The exponential decay time of the magnetic field is (a) $25$ mins [$82.6$ hours], (b) $71$ mins [$36.3$ hours], (c) $30$ mins [$24$ hours] for alive [dead] cockroach.
The average exponential decay time over all measurements is $50 \pm 28$ mins ($47.5 \pm 28.9$ hours) for alive (dead) cockroaches.
The offset magnetisation of 0.38 $\mu$G (thin dashed line) is attributed to the cockroach container dominating the signal for unmagnetised cockroaches.
Each data-point has an uncertainty of 0.08 $\mu$G.
Note different vertical scales in different panels.}}
\label{fig_decay}
\end{figure}

The general idea of our experiments is depicted in Fig.~\ref{fig_roach}.
Our setup and detailed methodology are described in the Methods section.
In short, magnetised cockroaches were measured with an all-optical Caesium atomic magnetometer with periodically moved sample tracking the temporal dependence of the magnetisation.
An exemplary set of measurement results is presented in Fig.~\ref{fig_decay}.
We conducted $15$ measurements, each lasted longer than $10$ hours: $8$ measurements on alive cockroaches and $7$ on the dead ones, see Supplementary Information (SI) for experimental data.
Additionally, more than $10$ shorter experiments were conducted ($2-5$ hours) confirming the trends described below, but are excluded from statistical analysis due to different experimental conditions.

All insects produced measurable magnetic fields.
Seven out of eight alive cockroaches gave rise to exponential magnetic field decay with an average decay time of $50 \pm 28$ minutes.
For one alive cockroach we observed a weak stable-in-time signal and this dataset is not included in the average calculation.
For another alive cockroach we observed magnetisation revival after the exponential decay, see SI.
All measured dead cockroaches gave rise to a stable in time magnetic field similar to data presented in Fig.~2.
The decay times have the average of $47.5 \pm 28.9$ hours.

\begin{figure}
\centering
\includegraphics[scale=0.08]{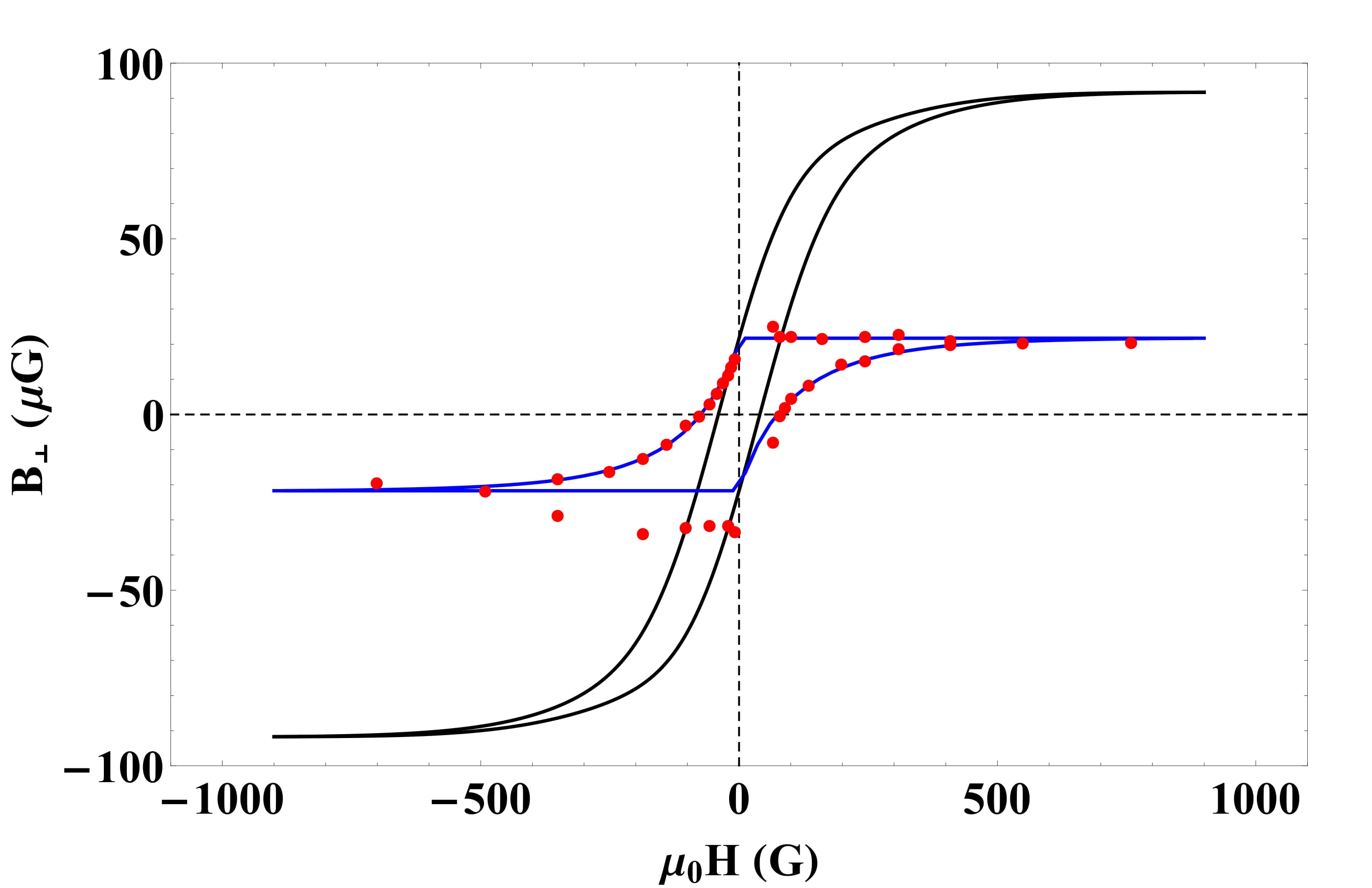} 
\caption{{\small Hysteresis measurements on dead cockroach.
The red dots show measured data of remanent magnetic field as a function of applied magnetic field.
The blue curve represents a fit to the data obtained with the software package HysterSoft~\cite{Dimian.2014} using the Preisach model.
The black curve is an exemplary hysteresis giving rise to our measured remanent hysteresis curve.}}
\label{fig_hysteresis}
\end{figure}

In order to gain further insight into the origin of the observed magnetism we conducted a remanent hysteresis measurement the results of which are presented in Fig.~\ref{fig_hysteresis}.
Note that this measurement is conducted on a single cockroach only.
It motivates our discussion on a possibility of greigite-based magneto-reception in cockroaches.

We model the observed decays with Brownian rotations of magnetic materials.
In the Methods section we describe a model that takes into account volume distribution of the magnetic deposits.
All its predictions are here explained in simple terms by assuming that observed magnetic relaxation curves match the exponential decay $M_0 \exp(- t / \tau) + $ offset,
where $M_0$ and $\tau$ are experimentally determined initial magnetic field and decay time respectively.
For Brownian motion of a spherical grain we have:
\begin{equation}
\tau = \frac{3 V \eta}{k_B T},
\label{EQ_TAU}
\end{equation}
where $V$ is the effective hydrodynamic volume of a deposit representing the whole sample, $\eta$ viscosity of its environment, $T$ denotes the environment's temperature and $k_B$ Boltzmann constant.

Since no change is expected in the volume of the magnetic materials inside alive and dead insect, 
Eq.~(\ref{EQ_TAU}) predicts that ratio of the decay times from dead and alive specimen is given by $\tau_d / \tau_a = \eta_d / \eta_a$, i.e. the ratio of respective viscosities.
Hence, the variation in the decay times from alive and dead cockroaches is explained by post-mortem increase in viscosity of the environment of the magnetic particles.
From measured data, the viscosity increment is by two orders of magnitude.

The data in Fig.~\ref{fig_decay} shows different values of the initial magnetic field of alive and dead insect that of course must also be explained by the model.
At the end of the magnetisation stage the magnetic field originating in the dead insect can only be smaller than the field generated at this stage by alive insect. 
There are two contributions diminishing the field:
(i) Since the cockroaches were washed in ultrasonic bath after death, some external magnetic materials could be cleaned;
(ii) The increased viscosity may cause partial alignment of internal magnetic deposits with the external field.
We first consider the case where the difference in the produced magnetic field is solely from contribution (i).
The ratio of the initial magnetic fields, $r = M_0(\textrm{dead}) / M_0(\textrm{alive})$, is now just another variable in the model that characterises the amount of magnetic materials washed out.
In this case experimental data does not uniquely specify all the parameters of the model, but only fixes the product $V \eta$ according to Eq.~(\ref{EQ_TAU}).
The second case is more interesting from a theoretical perspective. 
Assume that there were no external magnetic materials so that contribution (i) can be ignored.
In the Methods section we show that partial alignment in highly viscous environment is characterised by the alignment time which only depends on the viscosity, see Eq.~(\ref{EQ_ALIGN}).
In this case experimental data uniquely determines parameters of the model.
Their orders of magnitude can be estimated as follows. 
For the effect of partial alignment, the alignment time $t_{\otimes}$ in Eq.~(\ref{EQ_ALIGN}) has to be comparable with $20$ minutes as this is how long the cockroaches were magnetised.
Assuming they contain magnetic materials similar to magnetite or greigite, with saturation magnetisation $~10^5$ A/m, Eq.~(\ref{EQ_ALIGN}) predicts the viscosity $\eta_d \sim 10^7$ Pa sec.
Accordingly, the viscosity of alive insects is $\eta_a \sim 10^5$ Pa sec.
The hydrodynamic volume then follows from Eq.~(\ref{EQ_TAU}) and the estimated radius of magnetic deposits is $10 - 100$ nm.

Although magnetic field of dead insects at the end of the magnetisation stage is smaller than the magnetic field of alive insects, the ratio of initial magnetisations $r$ measured in our setup could be higher than $1$.
This is due to two-minute delay caused by loading the cockroach into magnetometer during which magnetisation of alive insect might already drop below the magnetisation of the dead specimen.
According to the model above one finds $r \lesssim 1 + 2/\tau_a$, where $\tau_a$ is in minutes.
Not all of our datasets satisfy this bound as clearly visible in Fig.~\ref{fig_decay}a.
This is an indication of underlying multimodal volume distribution.
Indeed, the curves in Fig.~\ref{fig_decay}a could be fit with a bimodal distribution with some particles exceeding $100$ nm radius, which intuitively can be understood as follows.
The small particles decay faster and this combined with the effect of the large viscosity difference leads to the big discrepancy between initial magnetisations of alive and dead insect
(taking into account that it takes $2$ mins to load the cockroach into the magnetometer).
However, this decay is too fast to match the experimental decay times and therefore larger particles have to be introduced.

\section{Discussion}

We showed that partial alignment of magnetic deposits requires a very high viscosity on the order $10^5$ Pa sec in alive cockroaches.
Quite high viscosity also follows in the case where some magnetic materials are washed out.
Assuming that cockroaches contain single-domain magnetic crystals of radius $50$ nm, 
experimental data shows that they would rotate in an environment of viscosity $\sim 10^2$ Pa sec in alive insect.
The obtained high values of viscosity agree with estimations for cytoskeleton inside cells, which was suggested to be a glassy material~\cite{Fabry.2001}.

In this context we note that cockroaches may also contain magnetic particles surrounded by a very low viscosity medium.
The time resolution of our experiment is set by the time it takes to load a cockroach into the magnetometer, about $2$ minutes.
Therefore, particles with a relaxation time on the order of tens of seconds are unnoticed. 
Eq.~(\ref{EQ_TAU}) then shows that in a low viscosity, water-like environment spherical particles up to a micron in radius are not detected.

The increment in viscosity of the environment inside dead cockroaches follows from the expected physiological changes.
The dead cells permanently dehydrate causing an increase in volume fraction of cytoskeleton which in turn increases environmental viscosity.
The increase in viscosity has already been independently observed inside dying cells~\cite{Kuimova.2009b}.

We note that in the case of partial alignment the estimated size of magnetic materials is in the range of biogenic clusters of magnetic particles that were extracted from two species of termites~\cite{Maher.1998b}.
This might not be a coincidence as termites are eusocial cockroaches~\cite{Inward.2007}. 
They belong to the same order Dictyoptera and have closely related symbionts.
Nevertheless, we emphasise that our experiments do not prove biogenic magnetism because environmental ferromagnetic contaminants could still be present in the tissues~\cite{Kobayashi.1995}.

The hysteresis measurement (see Fig.~\ref{fig_hysteresis}) reveals a small value of the coercive field, $40 \pm 6$ G. 
This is consistent with the measurement on termite \emph{Neocapritermes opacus}~\cite{Ferreira2005}.
The value is typical for multi-domain materials, ultrafine-grained particles of magnetite~\cite{Geiss2008} or single-domain greigite~\cite{DiazRicci.1992},
but not single-domain magnetite found in various other animals for which the corresponding coercive field is about $400$ G~\cite{Diebel.2000b}.
We pursue further the hypothesis of single-domain greigite.
Ref.~\cite{Cruden.1979b} identified a black region in the cockroach, \emph{Eublaberus posticus}, hindgut rich in metal sulphides.
Unknown organisms producing H$_2$S were also found in the flora around the black band.
Greigite could form there naturally by reduction of iron in H$_2$S rich environment with low oxygen level.
Since guts are well-nerved one can speculate that cockroaches could sense magnetic field based on greigite deposits connected to nerves in their hindgut.
See Refs.~\cite{Gould.1978, BEE_MAGNET, Liang.2016, Pan.2016, Lambinet.2017} for support of insect magneto-reception through abdomen.

Some evidence already exists for magneto-reception in cockroaches.
Cockroach mobility increases when it is placed in an Earth-strength magnetic field ($\sim 0.5$G) with periodically varying direction~\cite{Vacha.2006b} even after ablation of antennae~\cite{Vacha.2008b}.
This effect is absent under very interesting conditions:
(i) in total darkness~\cite{vacha.pnas}; 
(ii) when a radio-frequency field is added on top of the periodically varying Earth-strength magnetic field~\cite{Vacha.2009b};
(iii) in genetically modified cockroaches without Cry2 cryptochromes~\cite{vacha.pnas}.
Furthermore, the effect depends on light intensity and wavelength~\cite{vacha.pnas} providing 
support to the hypothesis that cockroaches utilise the so called radical-pair model as basis for their magneto-reception.
We now add another piece of evidence supporting this hypothesis.
A single-domain greigite particle of radius $50$ nm has the magnetic energy in the Earth's magnetic field almost $100$ times above the thermal energy at room temperature and hence it tends to align with the field.
However, its alignment time in Earth's magnetic field would be in the order of hours in a $10^5$ Pa sec viscosity environment (estimated under the assumption that no external magnetic contaminants are present on the measured cockroaches).
To the contrary, the period of the Earth-strength magnetic field variation in the described behavioural experiments is only $10$ minutes.

In conclusion, we realised a non-invasive method of magnetic field measurements applicable to biological systems at physiological temperatures.
The method employs an atomic magnetometer with periodically moved sample.
Our setup can also be used to infer other physical quantities than magnetic fields as illustrated with the determination of viscosity and volumes using the theory of Brownian motion (magnetorelaxometry).
When applied to American cockroaches the method clearly discriminates between dead and alive insects.
We propose that its physical origin lies in the post-mortem change of the viscous intracellular environment surrounding magnetic particles.
More generally, this technique can be used to test physiological processes or gain insights into the study of stochastic processes driving the motion of biomolecules or magnetic tracer substances inside cells.
Independent of any potential magneto-receptive suitability, the characterisation of physical properties of magnetic particles found in biological tissues of various species, even in the human brain~\cite{Kirschvink1992}, is worth studying. 
It may help to establish their physiological purpose, which is largely unknown, and to distinguish biogenic precipitates from magnetic contaminants. Our work also highlights the rarely addressed aspect of magnetisation dynamics, which poses limits on the usefulness of biomagnetic ferrites for navigational purposes. 
Static values of remanence and coercive field strength of magnetic particles are not by themselves sufficient to allow conclusions about magnetosensory suitability, the influence of their surrounding has to be taken into account.

\section{Methods}

\subsection{Experimental procedure and data collection}

Alive adult female and male cockroaches were kept in transparent insectaria with unlimited water and a diet consisting of cat food pallets and photoperiod of $12$light : $12$dark hours.
Before the experiment the insectarium was placed in a $4^{\circ}$C environment in order to immobilise the insects.
The experiments on dead cockroaches took place at least $2$ days after death and the dead cockroaches were kept in the $4^{\circ}$C environment to slow down putrefactive processes.
The death was induced by a nitrogen gas atmosphere and thereafter the cockroaches were washed in an ultrasonic bath confirming interior origin of the observed magnetic signal.
Every immobilised or dead cockroach was placed in a plastic bag which in turn was put in-between two characterised permanent magnetic plates producing a field of $3$ kG at the cockroach location.
The direction of induced dipole moment is indicated in Fig.~1.
After $20$ minutes of magnetisation the sample was mounted on a motorised translation stage inside the optical Caesium atomic magnetometer as shown in Fig.~4.
We periodically varied the distance between the sample and the Cs cell for at least $10$ hours.
One period takes $20$ seconds.
The recorded data-points are obtained by averaging over $20$ periods.
Each data-point has a magnetic field uncertainty of $0.08 \mu$G.
We verified that the bag alone produces a stable magnetic field over the timespan of our experiment ($0.38 \pm 0.08 \mu$G).

\begin{figure}
\centering
\includegraphics[scale=0.7]{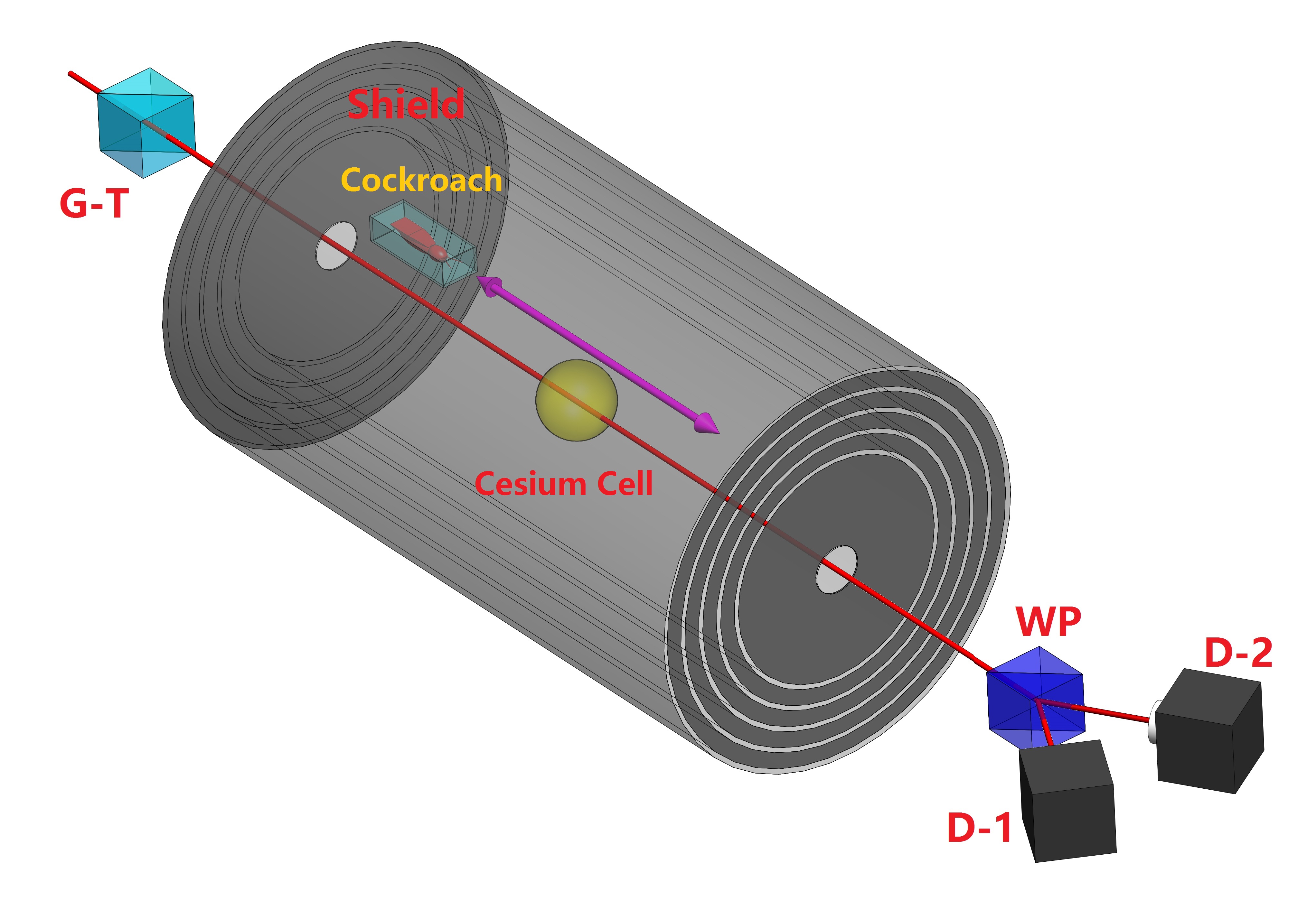} 
\caption{{\small Experimental setup. Only elements essential to the current experiment are shown.
See Ref.~\cite{Crepaz.2015} for the details for a similar setup.
Linear polarised light from a Glan-Thomson polariser (G-T) is coupled into a paraffin coated Cs cell.
The cell is surrounded by a $5$ layer magnetic shield to suppress ambient fields. 
After the light passes through the cell the induced polarisation rotation is recorded by a Wollaston prism (WP) and balanced detector assembly (D1 and D2).
This rotation carries information about the magnetic field at the Cs vapour.}}
\label{fig_setup}
\end{figure}

\subsection{Measurement of remanent hysteresis}

We placed a dead cockroach in between two N52 grade neodymium permanent magnet plates with the applied magnetic field oriented as in Fig.~1.
By changing the separation between the plates the magnitude of the magnetic field at the cockroach position was varied.
The magnetic field variation over the cockroach body is at most $15\%$.
The magnetisation time is $10$ minutes.
We start with an external field equal to $750$ G and continue along the hysteresis loop.

\subsection{Model}

\subsubsection{Relaxation}

Assume that the $i$-th magnetic particle has magnetic dipole moment $\vec \mu_i$.
Initially the cockroaches are magnetised along $z$ with macroscopic magnetic moment $M_z = N \langle \mu \rangle + A$, where $N$ is the total number of rotating particles, 
$\langle \mu \rangle$ is the mean dipole moment and $A$ is an offset value due to non-rotating magnetic materials.
Diffusion causes rotations of the magnetic deposits, and due to the random character of thermal forces components of the macroscopic moment orthogonal to $z$ vanish, $\langle M_x \rangle = \langle M_y \rangle = 0$.
The $z$ component decays as now the $i$-th rotated microscopic moment contributes with the projection $\mu_i \cos\theta_i$ along the $z$ axis. 
It is convenient to first consider the contribution to the average macroscopic moment from the particles with the same volume.
For Brownian rotations the average cosine at time $t$ is well known and given by the exponential decay: $\langle \cos \theta \rangle = \exp(- t / \tau)$, 
where the decay time, $\tau = 1/2D_r$, is the inverse of the doubled rotational diffusion coefficient.
For simplicity we assume spherically shaped rotating deposits and the decay time reads:
\begin{equation}
\tau_{V} = \frac{3 V \eta}{k_B T},
\label{TAU}
\end{equation}
where $V$ is the hydrodynamic volume of the deposit, $\eta$ viscosity of its environment, $T$ denotes the environment's temperature and $k_B$ Boltzmann constant.
Smaller magnetic particles give rise to a faster decay and they also produce a weaker magnetic field as $\mu_i = M_s V_i$, where $M_s$ is the saturation magnetisation.
The macroscopic moment is obtained by additional averaging over the volumes and reads $\langle M_z \rangle = N M_s \int f(V) V \exp(- t / \tau_V) \mathrm{d} V + A$, where $f(V)$ is the volume distribution.
We take the exponential distribution $f(V) = (1/\overline{V}) \exp(- V / \overline{V})$, with $\overline{V}$ being the average volume, and note that similar results are obtained with a log-normal volume distribution.
The average macroscopic moment can now be integrated to the closed form:
\begin{equation}
\langle M_z \rangle = \frac{2 N \bar \mu \, t}{\tau_{\overline{V}}}  K_2 \left( 2 \sqrt{t / \tau_{\overline{V}}} \right) + A,
\label{MZ}
\end{equation}
where $\bar \mu = M_s \overline{V}$ and $K_2(x)$ denotes the modified Bessel function of the second kind.
Fits of this model are presented in Fig.~\ref{fig_decay}.

\subsubsection{Alignment}

For the alignment process we treat particle's rotation as overdamped harmonic oscillator.
This is a suitable approximation in the highly viscous environment.
Thermal effects are neglected due to much higher magnetic interaction energy.
The angle between magnetic moment and the external field (counted from the field direction) changes as $\theta_t = \theta_0 \exp(- t /t_{\otimes})$, 
where $\theta_0$ is the initial angle and the alignment time $t_\otimes$ reads (see SI):
\begin{equation}
t_\otimes \simeq \frac{6 \eta}{M_s B}.
\label{EQ_ALIGN}
\end{equation}
Since unmagnetised cockroaches show no magnetic field, we assume a uniform in space distribution of initial magnetic moments.
It follows that at time $t$ the distribution of angles is still uniform, but in a smaller range $\theta_t \in [0,\theta_{\max}]$, where $\theta_{\max} = \pi \exp(- t /t_{\otimes})$.
This leads to the $z$ component of the macroscopic dipole moment being diminished by a factor of $\cos^2(\theta_{\max}/2)$ as compared to the value when all magnetic moments are aligned.
In our experiments, the cockroaches were magnetised for $20$ minutes and hence we use this number for $t$ in $\theta_{\max}$. 
The saturation magnetisation is chosen as $M_s = 3 \times 10^5$ A/m, with the order of magnitude matching magnetite and greigite.
After magnetisation it then takes $2$ minutes to mount the cockroach container in the magnetometer during which Brownian rotations cause partial demagnetisation according to Eq.~(\ref{MZ}).
At this stage we fit the model to the observed data.
We first fit measurement results from alive insect and then vary only viscosity to fit the data from dead cockroach.

\section{Acknowledgments}

We thank Li Yuan Ley for assistance in the initial setup of the magnetometer, the group of Antoine Weis, Universite de Fribourg, for the paraffin coated Cs vapor cells, Timothy Yeo for help with figure~\ref{fig_roach}.
{\sffamily Funding:}
This work was supported by DSO National Laboratories, Singapore and Centre for Quantum Technologies under project agreement DSOCL12111,
the National Research Foundation and the Ministry of Education of Singapore, start-up grant of the Nanyang Technological University and MoE grant no RG 127/14.
L.-J. K. gratefully acknowledges financial support from China Scholarship Council.

%%%%%%%%%%%%%%%%%%%%%%%%%%%%%%%%%%%%%%%%%%%%%%%%%%%%%%%%%%%%%%%%%%%%%%%
% SUPPLEMENTARY
%%%%%%%%%%%%%%%%%%%%%%%%%%%%%%%%%%%%%%%%%%%%%%%%%%%%%%%%%%%%%%%%%%%%%%%

\clearpage

\section*{Supplementary Information}

\subsection{Measured data}

Tables below list experimental data used in statistical analysis of the main text.
\begin{center}
\begin{tabular}{c | c | c | c }
 \multicolumn{4}{ c }{Alive}  \\ \hline
Dataset No & Cockroach No & Decay Time [min] & Initial Magnetic Field [Gauss]  \\
\hline \hline
1 & 1 & 25 & 7 \\ \hline % america
2 & 2 & 22 & 1 \\ \hline % cr1
3 & 3 & 89 & 0.54 \\ \hline % cr2
4 & 4 & 71 & 1.8 \\ \hline % cr3 mag1
5 & 4 & 77 & 1.1 \\ \hline % cr3 mag4
6 & 5 & 30 & 0.71 \\ \hline % cr4
7 & 6 & 37 & 0.71 % starved
\end{tabular}
\end{center}

\begin{center}
\begin{tabular}{c | c | c | c }
 \multicolumn{4}{ c }{Dead}  \\ \hline
Dataset No & Cockroach No & Decay Time [min] & Initial Magnetic Field [Gauss]  \\
\hline \hline
8 & 1 & 4956 & 17 \\ \hline % america
9 & 2 & 1374 & 0.77 \\ \hline % cr1
10 & 4 & 2178 & 1.2 \\ \hline % cr3 mag1
11 & 4 & 2178 & 1.05  \\ \hline % cr3 mag4
12 & 5 & 2136 & 0.5 \\ \hline % cr4 mag1
13 & 5 & 5676 & 0.8 \\ \hline % cr4 mag 2
14 & 5 & 1440 & 0.61 % cr 4 mag 3
\end{tabular}
\end{center}

\subsection{Fasted cockroach}
\label{APP_FAST}

We verified that the food pallets given to cockroaches can be magnetised and therefore conducted the experiment with fasted cockroach in order to exclude the hypothesis that the observed decay is due to ingested magnetism of the food.
The mean food transit time was determined to be $20.6$ hours, with a part of each meal retained in the crop for up to $4$ days~\cite{S.Snipes.1937}.
We therefore fasted the cockroach for $7$ days, giving it only water, and found that this had no effect on the decaying magnetic field.
This is not a conclusive proof of biogenic magnetism because environmental ferromagnetic contaminants could still be present in the tissues~\cite{S.Kobayashi.1995}.

\subsection{Magnetic field revival}
\label{APP_REVIVAL}

The theory outlined in the main text explains exponential magnetic field decay which was observed for most cockroaches.
For this one insect, however, we have seen a revival of the magnetic field as shown in the figure.

\begin{figure}[!h]
\includegraphics[scale=0.8]{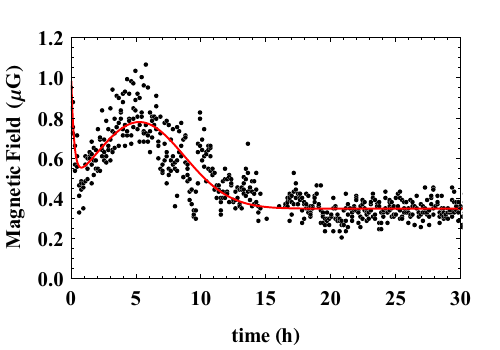}
\label{FIG_REVIVAL}
\end{figure}

\subsection{Alignment time}
\label{APP_ALIGN}

Consider a spherical particle endowed with magnetic moment $\vec \mu$ and surrounded by an environment with viscosity $\eta$ at room temperature $T$.
If the particle is subject to an external magnetic field $\vec B$, its rotational motion is described by the Newton law:
\begin{equation}
I \ddot \theta = - f \dot \theta - \mu B \sin\theta + \mathcal{T}.
\label{2LAW}
\end{equation}
Here $\theta$ denotes the angle between $\vec B$ and $\vec \mu$ (counted from $\vec B$), 
$I$ stands for the moment of inertia of the sphere, $I = \frac{2}{5} \rho V R^2$, 
and $f$ is the rotational friction coefficient, $f = 8 \pi \eta R^3$.
The next term gives magnetic torque and the last term is the thermal torque whose influence we will ignore here because the strong aligning field gives rise to $\mu B \gg kT$ even for very small magnetic moments. 
Despite its simplicity, for particles embedded in a highly viscous environment subjected to a constant field, similar models have successfully explained experimental results~\cite{S.eloi.2013}.

Our aim is to calculate the time it takes the particle to align with the field, $t_{\otimes}$.
Note that $t_{\otimes}$ is longer than the alignment time $t_{\downarrow}$ obtained when the magnetic torque is replaced by a stronger torque.
Similarly, $t_{\otimes}$ is shorter than the alignment time $t_{\uparrow}$ obtained if the magnetic torque is replaced by a weaker torque.
We show a simple upper and lower bound on the strength of the magnetic torque leading to alignment times that differ only by a constant factor of order one.
Hence the obtained formula also holds for $t_{\otimes}$.

Consider first the stronger torque $\mu B \sin\theta \le \mu B \theta$.
The original nonlinear problem now reduces to the damped harmonic oscillator.
Due to the assumed high viscosity $f^2 - 4 I \mu B \gg 0$, the oscillation is overdamped with the particular solution
\begin{equation}
\theta_t = \left( \theta_0 - \frac{r_- \theta_0}{r_- - r_+} \right) \exp(r_- t) + \frac{r_- \theta_0}{r_- - r_+} \exp(r_+ t),
\end{equation}
where
\begin{equation}
r_\pm = \frac{1}{2} \left(- \frac{f}{I} \pm \sqrt{\frac{f^2}{I^2} - 4 \frac{\mu B}{I}} \right).
\end{equation}
The initial conditions are: $\theta(0)=\theta_0$ for the angle and $\dot{\theta}(0)=0$ for the angular velocity.
Since $(f/I)^2 \gg 4 \mu B / I$ we simplify:
\begin{eqnarray}
r_+ & = & - \frac{\mu B}{f}, \\
 r_- & = & - \frac{f}{I} + \frac{\mu B}{f}.
\end{eqnarray}
Furthermore, both $r_{\pm}$ are negative with $r_- \ll r_+$ and therefore $\exp(r_- t)$ quickly decays to zero.
The long time dynamics is governed by the decay $\exp(r_+ t)$, which admits alignment time:
\begin{equation}
t_{\downarrow} = \frac{f}{\mu B} = \frac{6 \eta}{M_s B}.
\end{equation}
Note that this time is independent of the volume of the particle as well as its initial angle.

For the estimation of $t_{\uparrow}$ consider the magnetic moment at the initial angle $\theta_0$.
Since the torque tends to align it with the field, the accessible angles are from $\theta_0$ to $0$.
This gives rise to the following lower bound: $\mu B \frac{\sin\theta_0}{\theta_0} \theta \le \mu B \sin\theta$.
The problem reduces to the damped harmonic oscillator as above with the replacement $\mu B \to \mu B \frac{\sin\theta_0}{\theta_0}$.
Assuming uniform initial distribution of magnetic moments the average alignment time is:
\begin{equation*}
t_{\uparrow} = \frac{1}{4 \pi} \int_0^\pi \int_0^{2 \pi} d \varphi d \theta_0 \sin\theta_0 \frac{6 \eta \theta_0}{M_s B \sin\theta_0} = \frac{\pi^2}{4} \frac{6 \eta}{M_s B}.
\end{equation*}
Hence both upper and lower bounds on the average alignment time are size independent and of the same order of magnitude as $\frac{\pi^2}{4} \approx 2.5$.

\end{document}